# Geometrically Templated Dynamic Wrinkling from Suspended Poly(vinyl alcohol) Soap Films

*Yuchong Gao, Yinding Chi, Mohit Patel, Lishuai Jin, Jiaqi Liu, Pierre-Thomas Brun, and Shu Yang\**

Y. Gao, Y. Chi, M. Patel, L. Jin, J. Liu, S. Yang
Department of Materials Science and Engineering, University of Pennsylvania, 3231 Walnut Street, Philadelphia, PA 19104, United States
E-mail: shuyang@seas.upenn.edu

P.-T. Brun
Department of Chemical and Biological Engineering, Princeton University, Princeton, NJ 08544, United States

**Abstract:** Wrinkling is commonly observed as mechanical instability when a stiff thin film bound on a compliant thick substrate undergoes in-plane compression exceeding a threshold. Despite significant efforts to create a broad range of surface patterns via wrinkling, little has been studied about a dynamic and transient wrinkling process, where a suspended polymer thin film undergoes liquid-to-solid phase transitions. Here, a spontaneous wrinkling process is reported, when drying poly(vinyl alcohol) (PVA) soap films suspended on 3D printed wireframes with near zero or negative Gaussian curvatures. As water evaporates, a thickness gradient across the sample is developed, leading to non-uniform drying rates, and a concentration gradient between the inner and outer sides (exposed to air) of the suspended PVA soap film induces a differential osmotic pressure. Together, these effects contribute to an in-plane compressive stress, leading to the formation of surface wrinkles, whose growth is guided by the geometry of the frame. Importantly, the wrinkles evolve dynamically: the wavelength and number of the wrinkles can be tuned by altering the concentration of the PVA aqueous solutions, the initial mass, the relative humidity of the drying environment; the patterns of the resulting wrinkles can be programmed by the geometry of the wireframe.

# 1. Introduction

Wrinkles are surface patterns with undulated ridges and troughs, which are leveraged in a wide range of applications[1], including reversible adhesion[2,3], wetting[4,5], microfluidics[6], tunable optical devices[7,8], flexible electronics[9-11], solar cells[12], and cell alignment[13,14]. Wrinkles typically form as a result of compressive stress acting upon mechanically dissimilar bilayers. Examples include a thin stiff film of polystyrene or silica on top of a thicker and more compliant substrate such as polydimethylsiloxane (PDMS) or hydrogel. The thin film buckles to preserve its length and bends to minimize the elastic energy of the system when the compressive stress exceeds a critical threshold. These spontaneous periodic patterns typically emerge following a temperature change[15,16], solvent swelling[17], or in-plane mechanical compression[18-20]. Wrinkles of different wavelengths and amplitudes are generated depending on the intrinsic properties of the bilayer, their respective thickness, and the type of external stimuli [21]. Likewise, wrinkling can be driven by macroscopic confinement when suspending a thin rigid (tens to hundreds of nanometers thick) polymer film (e.g., polystyrene) on water[22-26]. There, capillary forces induce azimuthal compression, which is relieved by wrinkling throughout the film. When such a thin film is suddenly impacted by a steel sphere[25], the wrinkle wavelength can evolve dynamically as the impact draws materials radially inward. The patterns of surface wrinkles can be guided by the external force[23], unbalanced Gaussian curvatures induced by the cutout shape of the thin polymer shell[26], or top-down fabricated templates with nanometer-to-micron scale features[27-29]. Meanwhile, wrinkles can be reversibly formed and erased by exposure to light, heat, and external forces [30].

Despite significant efforts to create surface patterns of a wide range of morphologies and sizes via wrinkling instabilities, most studies come from bilayer systems where the wrinkles are formed in equilibrium. Under a constant input of stimuli, however, wrinkles can initiate and propagate, therefore, opening a pathway for creating the wrinkle patterns that evolve out of equilibrium. Dynamic wrinkle growth has been observed from the milk skin formed on water upon heating of the milk[24]. Treating the milk skin as a poroelastic thin film, Evans et al. attribute the evolving wrinkles to the evaporation-driven flow of water across the film, which imparts in-plane stresses; the time-dependent growth of wrinkles is determined by the viscosity of the complex fluid, milk, milk skin thickness, and the relative humidity. Wrinkling of a thin, partially suspended film is attributed to the residual stress when removing the substrate that the film is initially deposited on[31]. Curvature effect is also exploited to tune the wrinkling localization and amplitude on PDMS thin shells fixed on a silicone mat upon uniaxial stretching[32]. Nevertheless, dynamic wrinkling in a soft polymer film that is suspended on a rigid frame of different curvatures, undertaking the liquid-to-solid phase transition, has not been investigated, which will be important to enrich our ability to pattern the surface morphologies in a three-dimensional (3D) space[33,34].

Here, we dip coat the aqueous solutions of poly(vinyl alcohol) (PVA) on 3D printed wireframes of different geometric shapes and curvatures, creating freely suspended soap films. Compared with those obtained from the small molecule surfactants, which tend to burst during the liquid-to-solid phase transition[35], the PVA soap films created in our system do not burst during drying as a result of the long persistence length of the polymer chains, which have the weight-average molecular weight ($M_W$) in the range of 61,000 to 186,000 g/mol. As the PVA concentration, [PVA], increases during the drying process, PVA chains become increasingly glassy and entangled with each other, allowing the film to sustain chain slipping and thinning over time. Spontaneous wrinkling is observed during the water evaporation process and wrinkles' orientation is guided by the boundary confinements, that is the geometry of the wireframe. By varying the concentration of PVA solution, initial mass, and the relative humidity (RH) in the drying chamber, we can control the evaporation rate and the film shrinkage rate during the liquid-to-solid phase transition to induce unbalanced compressive stress across the film. We further investigate the effect of the molecular weight of PVA, the wireframe size, the diameter of the frame beam, and the surface chemistry of the wireframes to the dynamical tuning of the wrinkle wavelength and wrinkle numbers. The wrinkle generation process can be generalized to polygon and curved surfaces, forming periodic wrinkle patterns in previously unexplored geometries.

## 2. Results and Discussion

The PVA aqueous solution is dip coated on a 3D printed wireframe (**Figure 1A**), forming a suspended soap film with both sides exposed to air during the drying process. The entanglement of the long PVA chains provides the desired elasticity and stability to the soap film, which survives the liquid-to-solid phase transition without bursting. First, we fix the $M_W$ of PVA at 125,000 g/mol, mass of the initial PVA solution as 40 mg, the frame geometry and size (e.g., the side length of the equilateral-triangle-shaped frame $l_f$ = 20 mm), wireframe beam diameter $d_f$ = 2 mm, [PVA] =10 wt%, and RH =10% to study the wrinkle formation process (detailed fabrication and drying conditions shown in Figure 1B are summarized in Table S1). The thickness of the dried soap films is measured as ~ 10 µm at the center regions (Figure 1C), which increases towards the wireframe edges when the PVA solution forms the Plateau border with a negative line tension[36] (Figure 1D). During the drying process, wrinkles are spontaneously generated from the center where the solution is thinnest, and then evolve towards the edges, forming ordered wrinkles perpendicular to the vertices (Figure 1B).

To investigate the wrinkle generation and ordering during the liquid-to-solid phase transition, we first monitor the PVA soap film suspended on a triangular frame as a function of the drying time (see the setup in Figure S1). As seen in Figure 1B, the wrinkling process can be separated into three stages: initiation, propagation, and termination. At the beginning, the wet soap film has an average thickness ~ 480 ± 20 µm, and a thickness gradient forms spontaneously as the PVA solution pins on the wireframe. As water evaporates, the film shrinks, creating in-plane compressive stresses. Since the edges of the film are fixed by the wireframe, the internal stress cannot be fully relieved through contraction. To minimize the elastic energy associated with these stresses, the film undergoes out-of-plane deformation, forming wrinkles when the compressive stress exceeds the critical point. To understand the origin of the compressive stress, we introduce carbon black in the PVA solution as the tracer particles. As seen in Movie S1, the carbon black particles first flow to the edges of the wireframe, suggesting that PVA chains and water would also move outwards to the edges. This posit is confirmed by the scanning electron microscopy (SEM) images (Figure 1C-D), showing the formation of the Plateau border at the wireframe, while the center of the soap film is the thinnest (10.57 µm) vs. that at the edge (13.48 µm), indicating marginal accumulation of PVA at the edges during evaporation. Like in the case of the coffee ring effect, the evaporation rate is the highest at the edge. To keep the contact line fixed, water must flow outward from the center to the edge to replenish the depleted solvent[37,38], thus driving the movement of the colloidal particles to the meniscus at the edge. As the film center undergoes a liquid-to-solid phase transition (see the inset illustration of the cross-section of the film in Figure 1B at the initiation stage), residual stress starts to build up and grow towards the center, which is the thinnest. As the outer layers of the suspended film (exposed to air) continue to solidify and shrink, the inner layer remains largely unchanged if there is enough liquid. This effect creates an osmotic pressure difference $\Delta P_1$ between the inner layer and the outer layers along the film thickness, leading to isotropic compression at the center to initiate the global buckling[24]. As the center film becomes thinner and thinner, the drying front and the wrinkles dynamically propagate radial outward towards the edges (Figure 1B$_2$). Accordingly, drying-induced wrinkle patterns transit from the initiation regime to the propagation regime. The wrinkle pattern switches from the radial pattern to a vertex-guided pattern, where the wireframe boundary confines and regulates the direction of the propagating wrinkles (Figure 1B$_3$). At this stage, the motion of the PVA is switched from outward to inward, as manifested by the movement of the carbon black particles from the border to the center (see Movie S1). The wrinkles continue to increase in amplitude and wrinkle number with an enlarging elastic film front (Figures 1B$_4$ and 1B$_5$). Finally, the wrinkling process is terminated after the front reaches the wireframe and the final trace amount of water is evaporated. As the PVA film becomes increasingly glassified (that is more and more rigid) as it dries, the wrinkles formed on the soap film continue to evolve to minimize the strain energy, and eventually a glassy PVA film with guided surface wrinkles is formed (Figure 1B$_6$). To better elucidate the transient wrinkling process, we track the interactions between the PVA soap film and one edge of the square frame during the drying process (Figure S2, see Experimental Section for details), where the cohesive force applied on the surface of the PVA solution pulls the liquid inwards. Correspondingly, tension is applied on the edge of the frame initialized by the surface tension. The pulling force remains constant, as it is compensated by the solvent-evaporation-induced shrinkage, and thus wrinkling of the soap film.

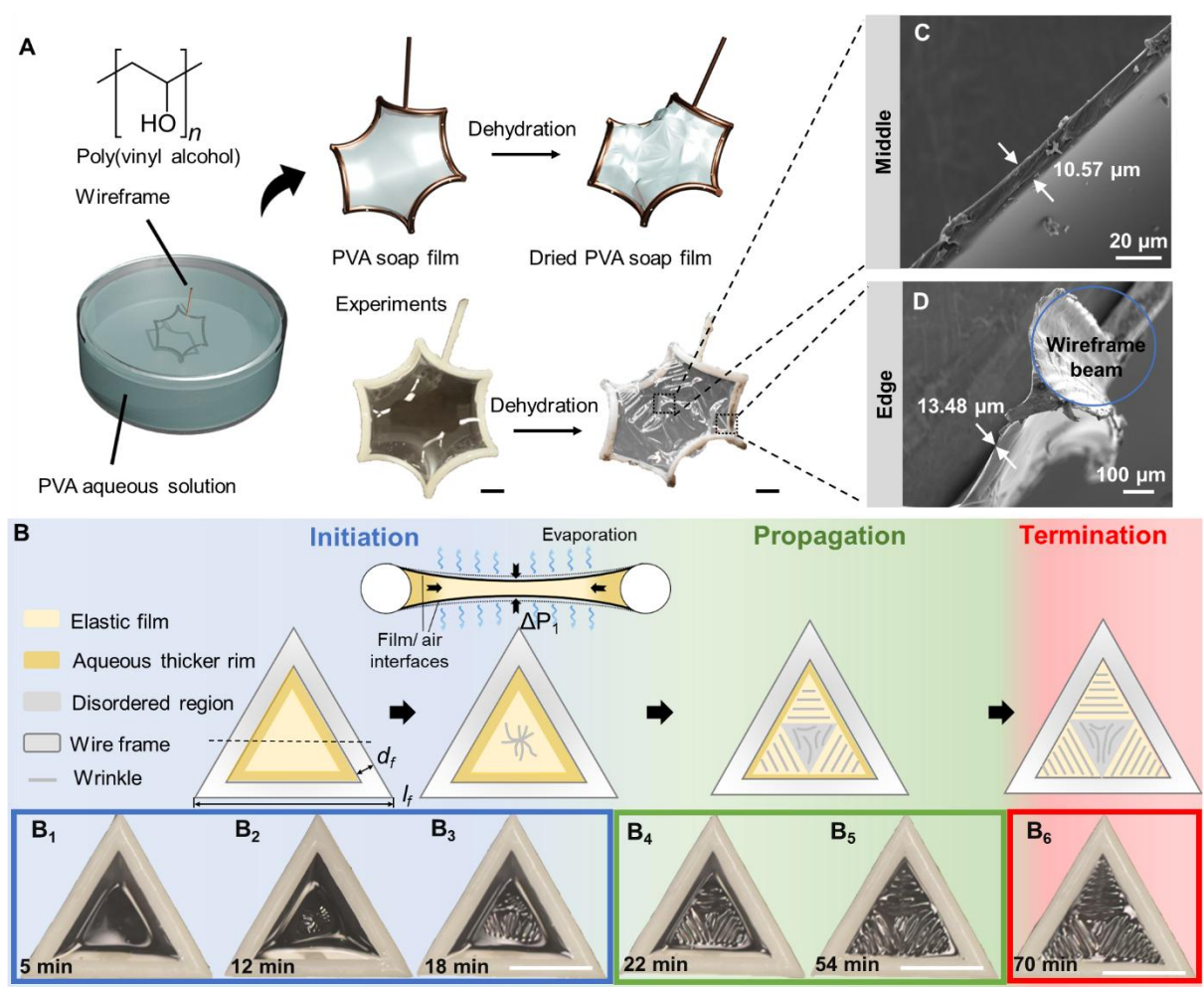

**Figure 1. Dynamic wrinkling in the PVA soap films during drying.** (A) Schematic illustrations of the dip-coating fabrication of a PVA soap film templated by a 3D printed wireframe and the corresponding photos from the experiments. Scale bars: 10 mm. (B)

Schematic illustration and the digital photos of a PVA soap film templated by a triangular wireframe dried at 23 °C, RH = 10% of variable drying time, displaying wrinkle generation and evolution in three stages: initiation, propagation, and termination. PVA $M_w$ =125,000 g/mol, PVA concentration is 10 wt%, initial coating mass is 40 mg, oxygen plasma treatment time is 5 min, the side length of the wireframe, $l_f$ = 20 mm, and the wireframe beam diameter $d_f$ = 2 mm. Initiation stage ($B_1$-$B_3$): formation of the initial elastic film in the center region at 5 min ($B_1$), radial wrinkle initiation in the center region at 12 min ($B_2$), and vertices guided transition at 18 min ($B_3$). Propagation stage ($B_4$-$B_5$): the beginning at 22 min ($B_4$) and the end at 54 min ($B_5$). Termination stage at 70 min ($B_6$). Scale bars: 10 mm, applicable to all Figure B panels. (C-D) Cross-sectional SEM images of the dried PVA soap film in the center region (C) and close to the wireframe (D).

To quantify the dynamic wrinkling process, we define and measure the wrinkle numbers $N$ and wavelength $\lambda$, the side length of the equilateral-triangle-shaped elastic region $l_e$ (i.e., the wrinkle propagation front), and the side length of the disordered wrinkled region in the center $l_d$ (see definition in **Figure 2A**). The wrinkling process is recorded at RH = 10%, and the average wrinkle number $\bar{N}$, the average wrinkle wavelength $\bar{\lambda}$, $l_e$, and $l_d$ are plotted as a function of time in Figure 2. Sharp increases in both $\bar{N}$ and $\bar{\lambda}$ are observed in the initiation regime (0-18 min in Figure 2B). As wrinkles propagate in the radial direction before reaching the Plateau region, $l_e$ continues to increase, and there is no such a triangle-shaped region with disordered wrinkles in the center in the initiation regime, that is $l_d = 0$ (Figure 2C).

To validate our hypothesis on wrinkle initiation, we perform a simplified finite element modeling (FEM) on the boundary-confined elastic sheet with varied thickness profiles (Figure S3). Orthotropic thermal expansion is used to qualitatively emulate the in-plane buckling of the elastic film before it is regulated by the frame. We note that a more detailed quantitative FEM simulation is challenging due to the complex transient nature of the drying soap film, where the [PVA] gradients across the film thickness and from the border to the center of the film continue to change during drying. We start with a circular frame to eliminate the boundary effect of the wireframe. The circular edge is fixed, and the cross-sectional profile of the soap film is estimated based on experimental observation. FEM simulation suggests that the film with a uniform thickness profile (2 mm, same as $d_f$) buckles out-of-plane in the form of a bulge starting from the perimeter of the circular boundary (Figure S3B), while the film with a non-uniform thickness profile (500 μm at the center, and 2 mm at the edge) generates radial wrinkles (Figure S3C), in a good agreement with the experimental observations (Figure S3A). Clearly, continuous thinning of the soap film and a thickness gradient from the plateau border regulated by the frame are essential for wrinkle initiation, leading to unbalanced stresses both in-plane (from the center to the edge) and out-of-plane (from the outer film-air interfaces to the inner film) exerted to the film.

As the stress built-up and wrinkling pattern continues to evolve radially with the evaporation time, $N$ and $\lambda$ increase in a linear fashion (up to 20 min, Figure 2B). Next, wrinkling enters the propagation regime, where the geometry of the boundaries starts to drive the orientation of the wrinkles. In the case of using a triangular frame with sharp corners as a template, the tension exerted on the elastic film by the boundaries can be decomposed to a component along the angle bisector and another perpendicular to it. Therefore, as opposed to random wrinkles formed from a circular frame, wrinkle patterns of regulated orientations are formed from polygon templates that have well-defined stress fields. The wrinkles align with the direction of the maximal principal stress, propagating along the median of the triangle to its vertex (inset of Figure 2A, FEM simulation). Stress along the angle bisector drives the evolution of $N$, $\lambda$, $l_e$ and $l_d$ as a function of drying time (Figures 2B and 2C), while that perpendicular to the bisector preferably aligns the wrinkles parallel to the tensile stress[39]. In this 'vertex-guided' wrinkling regime (20-58 min in Figures 2B and 2C), the elastic film region enlarges as wrinkles continue to propagate outwards, with graduate coarsening of $\lambda$ but a nearly constant average wrinkle number $\bar{N}$. At this propagation regime, the elastic front continues to reach the Plateau border with increasing $l_e$; both the regulated and disordered wrinkle regions can be observed at the center with increasing $l_d$ (Figure 2C, the light green region).

To investigate the mechanism that leads to the increase of $\lambda$ during wrinkle propagation, we vary PVA concentration, [PVA] (8% to 15%, Figure 2D, RH = 10%) and the initial coating mass (20 mg to 60 mg, Figure 2E, RH = 20%). The evolutions of $\bar{N}$, $\bar{\lambda}$, $l_e$, and $l_d$ are summarized in Figure S4 and S5, Movies S2-S4, and detailed fabrication conditions are summarized in Table S2 and S3. As shown in Figure 2D, the same scaling law of the wavelength coarsening vs. time, $\bar{\lambda} \propto t^{1/4}$ [24], is observed for films dried from different [PVA], although with different pre-factors. In addition, in the propagation regime (highlighted in light green in Figure S4), $\bar{\lambda}$ increases more significantly at higher [PVA]: at [PVA] = 5 wt%, $\bar{\lambda}$ increases from 492 μm to 622 μm; at [PVA] = 10 wt%, $\bar{\lambda}$ increases from 547 μm to 761 μm; and at [PVA] = 15 wt%, $\bar{\lambda}$ increases from 529 μm to 866 μm. This trend indicates that the intrinsic properties of the PVA solution, such as viscosity ($\eta$) could play an important role: as [PVA] increases, viscosity also increases, which will slow down the wrinkle coarsening process, resulting in a larger $\bar{\lambda}$. Furthermore, the ¼ power law matches well with the viscoelasticity-based wrinkle coarsening reported in the literature[24,40,41], where the stress is relaxed through the growth of the wrinkle wavelength. To study the thickness effect, we keep [PVA] = 10 wt% and RH = 20% but vary the initial mass (and thus the thickness) of the PVA solution from 20 mg (0.27 mm thick), (0.54 mm thick), to 60 mg (0.82 mm thick) (evolutions of $\bar{N}$, $\bar{\lambda}$, $l_e$, and $l_d$ are summarized in Figure S5, detailed fabrication conditions are summarized in Table S3). The mass is small enough that we could neglect the distortion of the film thickness due to the gravity effect. Despite different drying rates, the wrinkle coarsening lines collapse altogether with the same scaling (Figure 2E), suggesting that the effect of initial mass or thickness is negligible. Since the film is thin, here, we neglect the gravity effect. Increasing coating mass increases the soap film thickness with a thicker inner liquid layer, while the degree of drying of the outer layer is governed by polymer concentration and polymer chain length.

In the wrinkle termination regime (58-74 min in Figure 2B and Figure 2C, the red region), sharpening of $\bar{N}$ and $\bar{\lambda}$ are observed, while $l_e$ and $l_d$ are plateaued. Precisely predicting the final $\bar{N}$ and $\bar{\lambda}$ at equilibrium is challenging, which requires quantitative knowledge of the water evaporation rate, polymer chain relaxation behaviors, and liquid-to-solid phase transition process. Instead, we focus on the dynamic process and fit the change of the average wavelength over time with an exponential decay:

$$\bar{\lambda} = \overline{\lambda_0} - a \cdot \exp\left[{-(t-t_0)}/{\tau}\right] \quad (1)$$

Here, $\overline{\lambda_0}$ is the average initial wavelength, $t_0$ is the time when the decay starts, $a$ is the pre-factor determined by the counter-acting terms such as the chain entanglement as a function of polymer M$_W$ and concentration, and thus, the chain mobility, and $\tau$ is the decay constant related to the speed of the sharpening process. As displayed in Figure 2F, increasing thickness of the PVA soap film prolongs the exponential decay due to the increased $\tau$ (see light yellow, blue, and green regions with the initial coating mass increases). The thinner film reaches equilibrium faster due to the lower bending stiffness. The shortest wrinkle formation time is ~ 40 min with an initial coating mass of 20 mg. Eq.1 suggests that $a$ decreases with an increase of [PVA]; as chain entanglement increases, there should be less change of $\lambda$. However, our experiments show that even at [PVA] as high as 15 wt%, $\bar{\lambda}$ continues to change significantly in the termination regime (Figures 2D, 2F). This suggests that Eq. 1 does not fully capture the complexity of the wrinkle evolution at a higher PVA concentration, which is beyond chain entanglement. Further investigation is needed to develop a more quantitative model. Overall, the fully dried wrinkles can be coarsened as [PVA] and initial coating mass increase separately. When the [PVA] increases from 6% to 15%, $\bar{N}$ drops from 9 to 3, and $\bar{\lambda}$ increases from 600±37.33 μm to 1500±28 μm (Figure S6). When the initial coating mass of PVA increases from 15 mg to 145 mg, $\bar{N}$ drops sharply from 10 to 2, and $\bar{\lambda}$ increases from 527±10.74 μm to 1000±31.22 μm (Figure S7). In-situ measurement of wrinkle amplitude is challenging without disturbing the soap film or altering RH. Therefore, we measure the wrinkle amplitude after the PVA soap film is completely dried. The results show that under consistent drying conditions, wrinkle amplitude decreases with the increase of [PVA], while varying the initial mass of PVA had no significant effect on wrinkle amplitude after drying (Figure S8). Additionally, we vary RH while keeping other parameters constant, and the results will be discussed in detail later.

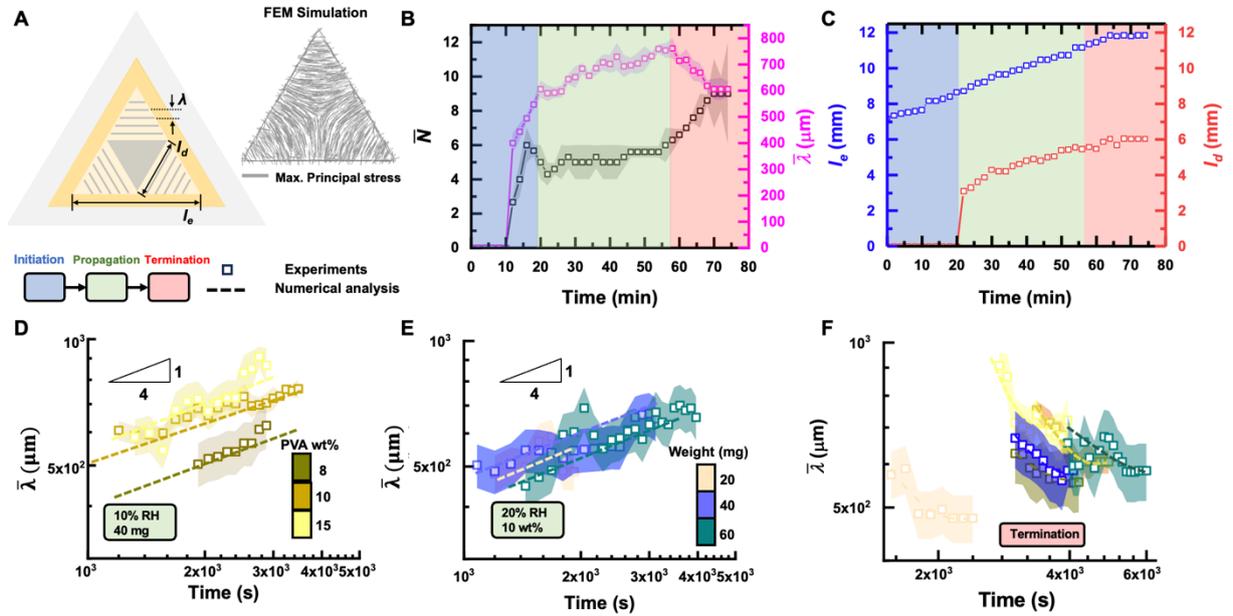

**Figure 2. Characterizations of the dynamic wrinkling process from drying PVA ($M_w$, 125,000 g/mol) solutions.** (A) Schematic illustration of the PVA soap film coated on a triangular frame at the propagation stage and the corresponding Finite Element Method (FEM) simulation result. The grey lines denote the orientations of the simulated maximum principal stresses in the PVA soap film. (B) The evolution of the average wrinkle number $\bar{N}$ and the average wavelength $\bar{\lambda}$ as a function of drying time, shows distinct three stages: initiation (blue), propagation (green), and termination (red). [PVA] = 10 wt%, initial coating mass is 40 mg, RH = 10%. (C) The evolution of the front of the elastic film (side length of the elastic region $l_e$) and the front of the center random region (side length of the disorder region, $l_d$) as a function of drying time. (D-E) Experimental results of the wrinkle propagation as the average wavelength $\bar{\lambda}$ vs. the drying time from PVA solutions of different concentrations (D)

and different initial masses (E). Initial coating mass is 40 mg, RH = 10% for varied [PVA] (D); [PVA]=10%, RH =20% for varied initial coating mass (E). (F) Experimental results of the wrinkles formed in the termination regime as the average wavelength $\bar{\lambda}$ vs. the drying time for varied experimental conditions (PVA concentration and initial coating mass). The symbols are the same as those in (D) and (E).

To further generalize the effect of the boundary on wrinkle orientation, we fabricate different polygon wireframes (**Figure 3**) with a fixed $l_d = 2$ mm and an initial coating thickness of ~500 µm. Like the triangular frame, sharp corners in polygons direct wrinkle orientation to be perpendicular to the angle bisector, leaving a frustrated center region with disordered wrinkles (highlighted in the grey regions seen in Figure 3A, D-G) where stresses imposed from different directions compete. As the angle of the vertices increases, from square (Figure 3A), pentagon (Figure 3D), hexagon (Figure 3E), heptagon (Figure 3F), to octagon (Figure 3G), the film in the center becomes less affected by the boundary and completely random wrinkles are observed from a circular frame (Figure 3H).

Next, we fabricate dumbbell-shaped frames connecting two identical circles (Figure 3I), triangles (Figure 3J), and squares (Figure 3K) by a rectangular frame. Besides the corners, the parallel frames between the dumbbell also exert tensile stress on the soap film, thus, directing wrinkles perpendicular to the frames as seen in Figure 3I-K. The wrinkle orientation within the dumbbell-shaped frames is influenced by the geometry of the ends and the resulting stress distribution. Figure 3I features circular ends, which lack directional edges, resulting in an isotropic stress field and a random wrinkle orientation in the central rectangle. In contrast, in Figures 3J and 3K, the angular trapezoidal and pentagonal ends provide directional constraints that guide the wrinkles to align along the long side of the connecting rectangle. The experimental results are corroborated by FEM simulation (Figure 3B and Figure S9), confirming that the wireframe-guided wrinkle orientation is induced by the tensile stress developed during the drying process. Here, the maximum and minimum principal stress distributions depict the tensile and compressive stresses within the film, respectively. At each point, the principal stresses are orthogonal to each other. The minimum principal stress direction corresponds to the orientation of the maximum compressive stress and thus indicates the direction of the wrinkle propagation during the drying process. As the compressive stresses accumulate, wrinkles form perpendicular to the maximum compression direction, following the orientation of the minimum principal stress. When curvature is introduced into the frames, the stresses exerted on the film are altered. For example, on a frame with a positive curvature (1/8 of a spherical surface), the fixed boundary attempts to deform the negatively curved soap film into a positively curved shape. As a result of this change in curvature, wrinkles in the center are aligned radially to accommodate the more compressive stress in the center (Figure 3L). On a negatively curved frame, however, the soap film is more stretchable compared to that on a flat frame due to the matching curvature between the soap film and the boundaries. Therefore, wrinkles perpendicular to the bisectors are extended more toward the center, shrinking the size of the randomly wrinkled region (Figure 3M-N).

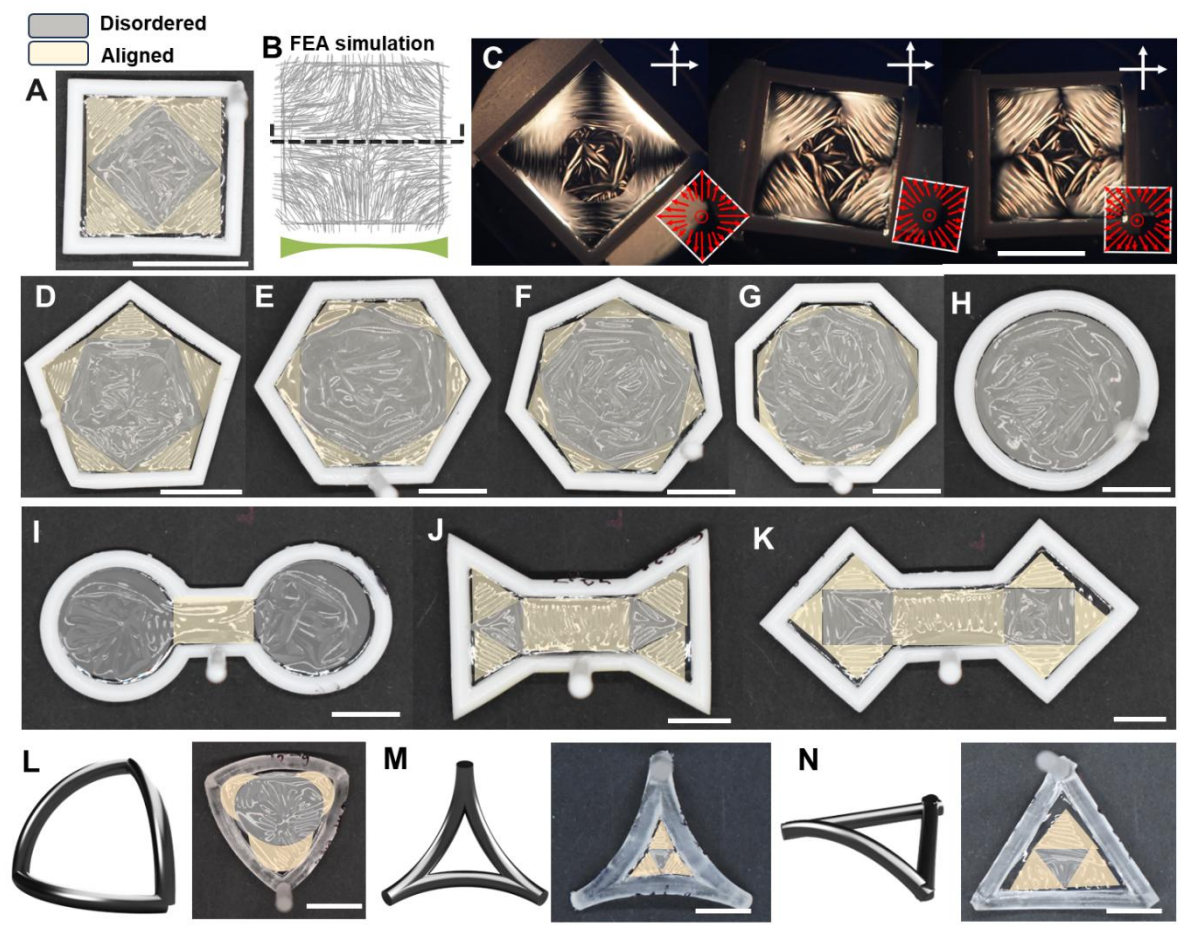

**Figure 3. Wrinkles formed on the dried PVA soap films templated by wireframes of various geometries, RH = 20%.** (A) Digital image of the PVA film templated from a flat square frame. Colors are added to indicate the disordered (grey) and the ordered (light yellow) wrinkles. (B) FEM simulation result, the grey lines denote the orientations of the simulated maximum principal stresses in the PVA soap film. (C) POM images show varying birefringence of a dried PVA soap film when the angle between the templated wrinkled film and the two cross polarizers are 45°, 15°, and 0°, respectively. (D-K) Flat wireframes with various geometries, including pentagon (D), hexagon (E), heptagon (F), octagon (G), circular (H), dumbbell connected circles (I), dumbbell connected triangles (J), and dumbbell connected squares (K). (L-N) Curved frames with positive (L) and negative (M and N) curvatures. Scale bars: 10 mm.

To visualize the evolution of the alignment within the wrinkling PVA films, we custom-build a polarized optical microscopy (POM) imaging setup, consisting of a white light source, a polarizer with slow axis perpendicular to the soap film plane, a rotating sample holder, an analyzer with slow axis parallel to the ground and a camera placed on the same optical path in sequence (Figure S10). The change of birefringence of the wrinkled PVA soap film as the sample rotates relative to the crossed polarizers (Figure 3C), suggesting anisotropic alignment of PVA chains in the wrinkled regions due to the local compression and the stress release process. The center region with random wrinkles shows no birefringence and remains dark regardless of the sample rotation angle, confirming the isotropic arrangement of the PVA chains.

Next, we explore the effect of the environmental RH on wrinkle formation as it affects the water evaporation rate and stress buildup. When RH increases from 10% to 90%, the evaporation rate slows down, and wrinkles are flattened (**Figure 4A**). Accordingly, the [PVA] gradient at high RH is reduced and a decrease in $\bar{N}$ is observed (Figure 4B), while $\bar{\lambda}$ first increases when RH increases from 10% to 50% but drops to zero as the wrinkles are flattened when RH ≥ 75%. Correspondingly, the amplitude of wrinkles becomes smaller with increasing RH (Figure S8C). Consistent with this, as RH increases from 10% to 60%, both $l_d$ and $l_e$ slightly decrease due to the coarsening of the wrinkle morphology. When RH reaches 75% or higher, however, the soap film becomes smooth, $l_e$ reaches maximum and $l_d$ is reduced to zero (as shown in Figure 4C). Comparing the complete wrinkling processes at RH = 20% (Figure 4A, ii and Figure S11) and 50% (Figure S12), the latter has a more regular center region (Figure 4A, iii), which can be ascribed to the much longer relaxation time enabled by the slow evaporation rate so that the frustrated random patterns in the center can be released, resulting in a flat center region. At RH > 75%, the [PVA] concentration gradient is too small to induce a large enough stress that can sustain the wrinkle formation on the fully dried soap film (Figure 4A, iv-v). Noticeably, at RH = 75%, wrinkles with small wavelengths are generated when the film is dried for 150 min due to the sudden increase of the stress when the drying front touches the boundary. The prolonged drying process and high water content at RH 75% give the PVA soap film enough time to relax back to a flat film; the wrinkles are eventually smoothened out (Figure 4D, Figure S13, Movie S5). When RH is increased to 90%, no wrinkles form (Figure 4A, v, Figure S14, Movie S6). The distinction between the initiation, propagation, and termination regimes becomes fuzzy when RH ≥ 75%. The wrinkle formation is significantly delayed, with most changes occurring in the termination regime when the film solidifies. Consequently, in Figure 4D, the wrinkle evolution is represented without delineating the three regimes, as they are not distinctly observable under these conditions, as shown in Figures 4D-4F.

Moreover, the drying speed can be tuned by the environmental RH as a pathway to dynamically control the wrinkle patterns. For example, the film can be first dried at RH 50% for 20 min (the initiation regime) to generate wrinkles, followed by moving it to a chamber at RH 90% for 300 min (Figure 4G, i). The generated wrinkles disappear. If the transfer happens in the propagation regime (RH 50% for 30 min), further drying at RH 90% does not eliminate all the wrinkles, but rather leaves a few short ones preserved as shown in Figure 4G, ii. If the transfer occurs in the termination regime (RH 50% for 60 min), the wrinkles with decreased $\bar{N}$ are observed and the random wrinkles in the center region are eliminated (Figure 4G, iii). By tuning the wrinkle patterns via dynamically changing the drying conditions, we shed light on creating a more complex library of wrinkle patterns. However, when the wrinkled PVA soap film is fully dried (e.g., as shown in Figure 4G, iii) and transferred to an environment with RH 20% for 48 h, the wrinkles cannot continue to develop or change (as shown in Figure S15A). Similarly, when a fully dried, wrinkled PVA soap film formed at RH 20% is transferred to an environment with RH 90% for 48 h, the pre-existing wrinkles cannot be erased (see Figure S15B). These results confirm that the wavelength and amplitude of the wrinkles are fixed after the PVA soap film is fully dried.

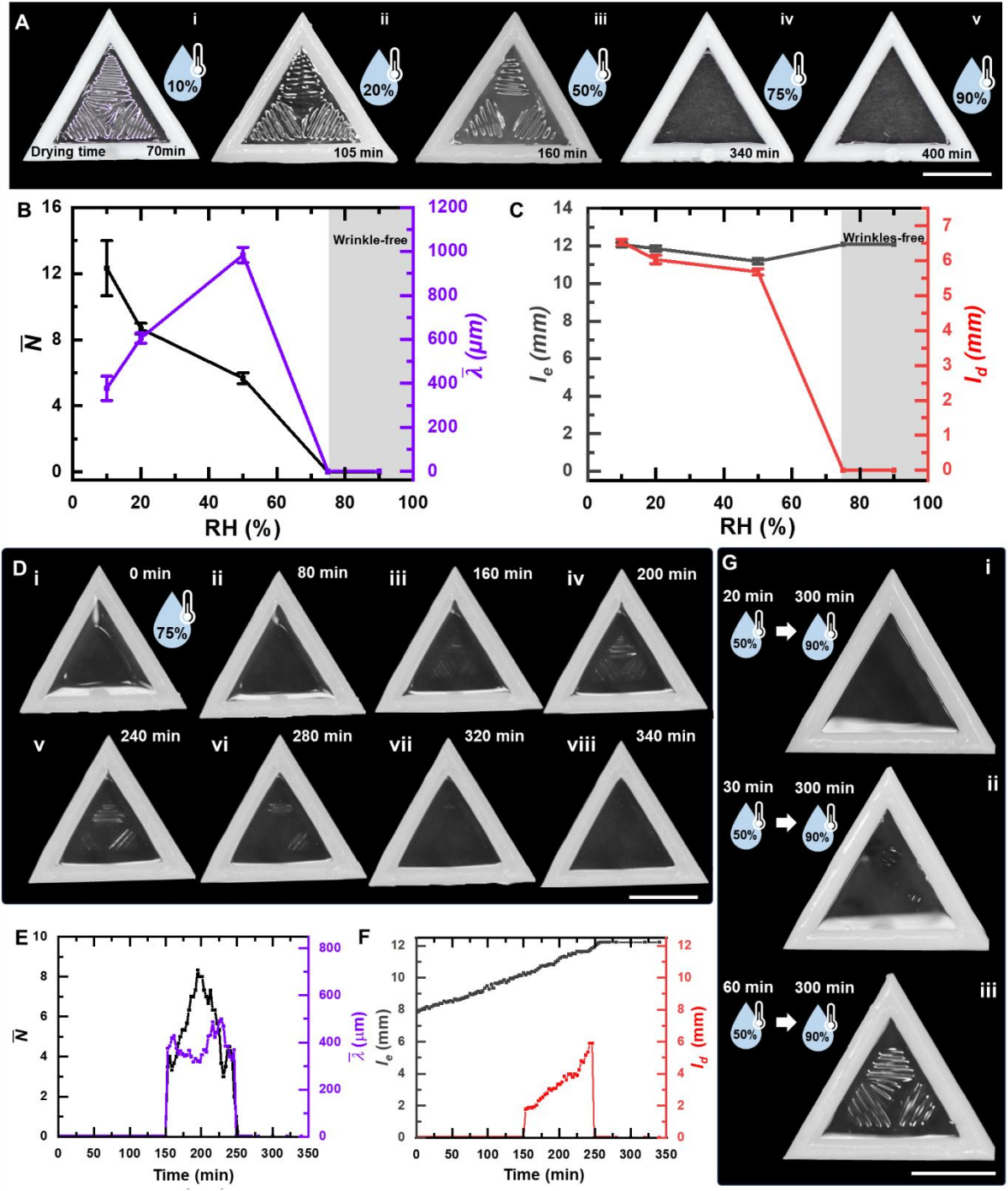

**Figure 4. Tunable wrinkle patterns by varying the environmental relative humidity (RH).** (A) Tunable wrinkle patterns on identical flat triangular frames (side length of the wireframe $l_f$ = 20 mm and the wireframe beam diameter $d_f$ = 2 mm) by varying RH from 10% to 90%. The time when the films are fully dried is recorded for varying RHs. (B) Evolution of the average wrinkle number $\bar{N}$ and the average wavelength $\bar{\lambda}$ as a function of RH. The film is wrinkle-free when RH ≥ 75%. (C) Evolution of the front of the elastic film (side length of the elastic region $l_e$) and the front of the center random region (side length of the disorder region, $l_d$) as a function

of RH. (D) The revolution of the wrinkle patterns of a PVA soap film (Mw, 125,000 g/mol) at different drying time when RH = 75%, templated by a triangular wireframe at 23 °C. PVA concentration is 10 wt%, initial coating mass is 40 mg, oxygen plasma treatment time is 5 min, the side length of the frame is 20 mm, and the wireframe beam diameter is 2 mm. The wrinkles appear at 150 min and disappear at 250 min. (E) Evolution of the average wrinkle number $\bar{N}$ and the average wavelength $\bar{\lambda}$ as a function of drying time when RH = 75%. (F) Evolution of the front of the elastic film (side length of the elastic region $l_e$) and the front of the center random region (side length of the disorder region, $l_d$) as a function of drying time when RH = 75%. (G) Wrinkle patterns generated at different environment RH: (i) 50% RH for 20 min, followed by moving to 90% RH for 300 min, (ii) 50% RH for 30 min, followed by moving to 90% RH for 300 min, and (iii) 50% RH for 60 min, followed by moving to 90% RH for 300 min. Scale bars: 10 mm.

We further vary $M_W$ of PVA, [PVA], initial mass, $l_f$, $d_f$ (the fully dried wrinkle pattern as shown in **Figure 5A-D**). At a fixed [PVA] =10 wt%, chain entanglement from the longer polymer chains ($M_W$, 61,000 g/mol to 189,000 g/mol) can slow down the propagation process, varying $\bar{N}$, without changing $\bar{\lambda}$ at 600 µm (Figure 5A, see detailed discussion in Supporting Information, Table S4, Figure S16). By varying [PVA] (Figure 5B), initial mass (Figure 5C), $l_f$ from 10 mm to 30 mm (Figure 5D, Table S5, Figure S17), $d_f$ from 1 mm to 5 mm (Table S6, Figure S18), oxygen plasma treatment time from 0 min to 25 min (Table S7, Figure S19), we measure the wrinkle area, size of the random region, $\bar{N}$ and $\bar{\lambda}$ obtained from the dried PVA soap films. At fixed parameters ([PVA]=10 wt%, $M_W$=125,000 g/mol, ~40 mg initial coating mass) and environmental conditions (22 ± 1°C, RH 15 ± 1.5%), all the wrinkles generated on the triangular wireframes have $\bar{\lambda}$ = 604.63 ± 63.88 µm. Increasing $l_f$ enlarges the surface area, leading to an increase of $\bar{N}$ and $\bar{\lambda}$, but does not alter $l_e$ and $l_d$ normalized by $l_f$ (Figure S17). The maximum wrinkle area, approximately 389.71 mm², is achieved at $l_f$ = 30 mm with an initial PVA mass of 90 mg. Likewise, variations in $d_f$ have a limited impact on the final wrinkle morphology (Figure S18). While oxygen plasma treatment enhances surface wettability, it does not significantly alter the drying dynamics or the stress distribution (Figure S19).

Lastly, we extend the wrinkling process using more complex wireframes with minimal surfaces, which are characterized by zero mean curvature, epitomizing geometric and physical minimalism to reduce material usage and generate efficient architectural forms. It appears that more regular patterns are formed on the curved frames compared to those formed on flat frames, leading to more guided wrinkle arrangements. For example, the helicoid gives rise to regular radial wrinkles on the surface due to the radial tension from the center wire and outside helix (Figure 5E), a unit of a Schwarz-P surface brings about a more regular version of wrinkle patterns compared with those generated from a flat hexagonal frame (Figure 5F), and a catenoid surface provokes vertical wrinkles around the whole surface from the tension exerts by the two circular frames (Figure 5G).

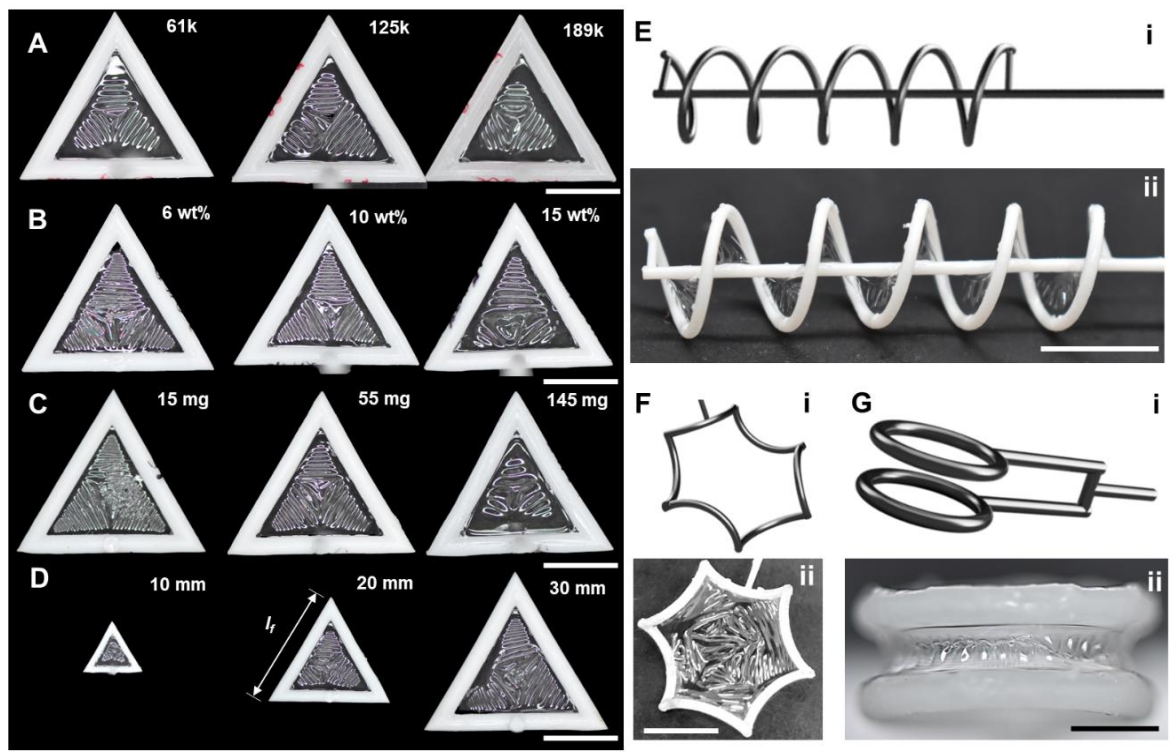

**Figure 5. Library of accessible wrinkle patterns via tuning the PVA properties and the boundary conditions.** (A-D) Tunable wrinkle patterns on identical flat triangular frames, compared to the pattern generated from a set of fixed parameters as the oxygen plasma treatment time (5 min), by varying PVA molecular weight $M_W$ (A), the PVA concentration (B), the initial coating mass (C), and the side length of the wireframe $l_f$ (D). (E-G) Wrinkles are generated on curved PVA soap films coated on a helicoid frame (E), a unit of the Schwarz-P surface (F), and a catenoid surface (G). Scale bars: 10 mm

## 3. Conclusion

In conclusion, we investigate a quasi-dynamic wrinkling process during the liquid-to-solid phase transition of PVA soap films suspended on 3D printed wireframes of different geometries. Initial Plateau border thickness gradient and the corresponding non-uniform drying rate (in-plane), and concentration gradient and thus differential osmotic pressure within the film (out-of-the-plane) during the liquid-to-solid phase transition of the PVA soap film are the key factors that contribute to the wrinkling. The average wrinkle number $\bar{N}$ and the average wavelength $\bar{\lambda}$ can be tuned by varying one or more parameters, including the frame side length $l_f$, the diameter of frame beam $d_f$, concentration and initial mass of the PVA solutions, PVA molecular weight $M_W$, and environmental RH. Unlike the traditional wrinkling on a rigid-soft bilayer or a thin rigid film suspended on water, our wrinkling mechanism can be extended to a wide range of suspended soap films consisting of polymers and their nanocomposites with inorganic nanoparticles, and on both flat and curved surfaces. Our findings also highlight the complexity of wrinkle formation, which is influenced by multiple parameters simultaneously. Such complexity makes it challenging to experimentally decouple the effects of viscosity of the PVA solution, the elasticity of the PVA film, temperature, and RH to generate a universal scaling law of wrinkling during the liquid-to-solid phase transition. We hope this work will inspire further investigation of this interesting and complex problem, and explore potential applications in flexible electronics, optical devices, and tissue engineering, as well as serving as templates to guide the assemblies of functional nanomaterials in complex 2D and 3D geometries.

## 4. Experimental Section

*Materials*: Poly(vinyl alcohol) (PVA) with weight-average molecular weights ($M_W$) of 23,000, 61,000, 89,000, 125,000, and 186,000 g/mol, and carbon black powders were purchased from Sigma-Aldrich. Sodium chloride (ACS reagent >99%), lithium chloride (ACS reagent >99%), potassium carbonate (reagent grade >98%), sodium sulfate (ACS reagent >99%, anhydrous, granular) were purchased from Sigma-Aldrich. Isopropyl alcohol (IPA) (70% v/v) and sulfuric acid (95%-98% w/w certified ACS plus) was purchased from Fisher Scientific. 3D printing UV curable White Model Resin was purchased from HARZ Labs LLC.

*3D Printing of the Wireframes*: Both the flat and curved wireframes were 3D printed using a Phrozen Mini 4K digital light processing (DLP) 3D printer and UV curable methacrylate monomer and oligomer-based HARZ Labs white model resin. The 3D models for printing were sliced to 50 µm layers in Chitubox software, the curing time for each layer was set to 8 s with 20 s light-off delay. After each layer was cured, the build plate was lifted up for 5 mm at a speed of 65 mm/min before returning to the position to cure the next layer. Immediately after the print, the build plate attached to the printed model was immersed in an isopropyl alcohol (IPA) bath and sonicated for 10 min in a Branson 3800 Ultrasonic cleaner to remove uncured resin, followed by air drying and 405 nm flood UV exposure to completely crosslink the resin.

*Preparation of the PVA Soap Films*: The PVA aqueous solutions of different $M_W$ and concentrations were prepared in deionized (DI) water. The mixture was magnetically stirred in a beaker placed in an 80 ℃ oil bath on a hotplate (IKA C-MAG HS 7) till it became homogeneous. The solution was completely cooled down to room temperature before dip coating. 3D printed wireframes were treated with oxygen plasma (Harrick Scientific Corp PDC-001 plasma cleaner/sterilizer) to enhance the affinity with the PVA solution and the stability of the soap film. The mass of the empty frame was recorded using a balance (Sartorius Entris224I-1S) before dipping into the PVA solution. After dipping, the excess PVA solution on the wireframe was wiped off. Since our PVA-based soap film is much more viscous than a typical surfactant-based soap film, it allowed us to gently add or remove small PVA droplets using a clean tweezer, followed by weighing repeatedly until reaching the desired weight. Then, the wireframe was immediately fixed on a polystyrene foam with the soap film fully suspended in air, which was placed in a closed, home-built glass chamber with controlled RH for drying and recording the wrinkle evolution. The RH inside the chamber (10-90%) was controlled using an open glass petri dish filled with sulfuric acid or aqueous saturated salt solutions. A closed chamber with concentrated sulfuric acid (98 wt%) and saturated LiCl aqueous solution had RH of 10 % and 20%, respectively, and that from saturated $K_2CO_3$, NaCl, and $Na_2SO_4$ aqueous solutions had RH of 50%, 75%, and 90%, respectively. The RH in the chamber was allowed to stabilize for at least 12 hours before use and was continuously recorded by a humidity monitor (Habor HM118A) during the soap film drying. After drying, the mass of the dried soap film with the wireframe was recorded. 0.01 wt% carbon black was well mixed with PVA aqueous solution (PVA molecular weight: 125000 g/mol) and soap film coated on the 3D printed triangular frame following the same process as the previous soap film coating steps.

*Characterizations*: Digital photos of the wrinkle formation process at different RH were taken by a Nikon D5600 DSLR camera at 180 s time intervals, the digital images of the wrinkle formation process at the RH = 20% with the carbon black added to track the flow direction and shrinkage of the wrinkled soap film were taken at 30 s time interval. All images were processed in ImageJ software to measure the average characteristic wavelengths. The average thickness of the wet soap films and the average amplitude of the wrinkles generated in the dry PVA soap films were measured using a caliper (Absolute Digimatic) placed perpendicular to the films and averaged over six locations on the same wrinkled pattern. Cross-sectional Scanning Electron Microscopy (SEM) images were obtained from FEI Quanta 600 Environmental Scanning Electron Microscopy (ESEM) at 10 kV electron beam under 0.50 Torr vacuum to measure the average dry film thickness at the center and the edge (see Figure 1A) using ImageJ. The macroscopic polarized optical images (POM) were recorded on a home-built set-up (see Fig. S7), including a white light source (Motic MLC-150C fiber optic illuminator), a linear polarizer (Thermo Fisher Scientific), a rotatable sample holder (Thermo Fisher Scientific), a rotatable waveplate holder (Thermo Fisher Scientific LCC1223-A), an analyzer with optical axis perpendicular to that of the polarizer (Thermo Fisher Scientific), and a digital camera, aligned on the same optical path.

*Finite Element Method (FEM) Simulation*: FEM simulations were performed in Abaqus/Standard (Dassault Systèmes) to simulate the stress developing process during the solid-to-liquid phase transition for templated PVA films. The geometry of different shaped films was imported into Abaqus CAE as a step file and meshed with 4-node tetrahedral elements (C3D4). A mesh refinement study was applied to verify the accuracy of the meshes. The mechanical properties of PVA were simplified and modeled using a linear elastic model assuming a constant Young's modulus of 50 kPa and a Poisson's ratio of 0.3 to qualitatively capture the low stiffness of PVA soap film in the initiation regime, which contains a large amount of water. The orthotropic coefficient of thermal expansion was set to 0.2/ °C in the in-plane direction and -0.05/ °C along the thickness direction to simulate the initial buckling through the expansion of the linear elastic region of the PVA film before the initiation stage during the drying process. The orthotropic coefficient of thermal expansion was set to -0.1/ °C in the in-plane direction, and -0.05/ °C along the thickness direction to simulate the stress conditions for the ordered wrinkle patterns regulated by the geometric templates. The edges were fully constrained to simulate the boundary constraints from the wireframes. Static simulation was performed as the temperature was raised to simulate the expansion process. The orientations of the principal stresses (aligned with grey lines) are plotted to indicate the direction of the propagation of the ordered wrinkles aligned with the direction of the minimum principal stress.

*Measurement of the pulling force on the frame*: We created a soap film from a PVA aqueous solution in a U-shaped frame (side length $l_f$ = 20 mm, beam diameter $d_f$ = 2 mm) with a slider (no friction between the frame and slider). $M_W$ of PVA was fixed at 125,000 g/mol, the mass of the initial PVA solution was kept at 25 mg or 40 mg, respectively, PVA concentration (10 wt%), and RH (43%) were maintained during the drying process. The slider was pulled away from the fixed side with a distance of 7 mm and constrained by a foam base. The force sensor coupled with Instron 5564 was connected to the frame to measure the force applied on the slider during the drying process of the PVA soap film.

**Acknowledgments**

S.Y. and P.-T.B acknowledge the support from National Science Foundation (NSF) Future Eco Manufacturing Research Grant (FMRG), # CMMI 2037097. The authors acknowledge the use of SEM instrumentation supported by NSF through the Laboratory for Research on the Structure of Matter (LRSM), University of Pennsylvania's Materials Research Science and Engineering Center (MRSEC) (DMR-2309043). The authors thank Randall Kamien for suggesting the use of 3D printed frames, and Eleni Katifori, Ian Tobasco, Anderj Košmrlj, and Baohong Chen for helpful discussions on wrinkle formation mechanisms.

**Author Contributions**

Y.G., Y.C., and S.Y. conceived the idea and designed the experiments. Y.G., Y.C., M.P., and J.L. performed the experiments, Y.C. and L.J. performed the simulation, Y.G. and Y.C. analyzed the data, P.-T.B. assisted with data analysis. Y.G. Y.C., and S.Y. wrote the manuscript. Y.G. and Y.C. contributed equally to this work.

**Conflict of Interest**

The Authors declare no conflict of interest.

# References

1. Chen, C.-M. & Yang, S. Wrinkling instabilities in polymer films and their applications. *Polym. Int.* **61**, 1041-1047 (2012). https://doi.org/10.1002/pi.4223
2. Chan, E. P., Smith, E. J., Hayward, R. C. & Crosby, A. J. Surface wrinkles for smart adhesion. *Adv. Mater.* **20**, 711-+ (2008). https://doi.org/10.1002/adma.200701530
3. Lin, P. C., Vajpayee, S., Jagota, A., Hui, C. Y. & Yang, S. Mechanically tunable dry adhesive from wrinkled elastomers. *Soft Matter* **4**, 1830-1835 (2008). https://doi.org/10.1039/b802848f
4. Lin, P. C. & Yang, S. Mechanically switchable wetting on wrinkled elastomers with dual-scale roughness. *Soft Matter* **5**, 1011-1018 (2009). https://doi.org/10.1039/b814145b
5. Xu, L. *et al.* Earthworm-Inspired Ultradurable Superhydrophobic Fabrics from Adaptive Wrinkled Skin. *ACS Appl. Mater. Interfaces* **13**, 6758-6766 (2021). https://doi.org/10.1021/acsami.0c18528
6. Khare, K., Zhou, J. & Yang, S. Tunable Open-Channel Microfluidics on Soft Poly(dimethylsiloxane) (PDMS) Substrates with Sinusoidal Grooves. *Langmuir* **25**, 12794-12799 (2009). https://doi.org/10.1021/la901736n
7. Chan, E. P. & Crosby, A. J. Fabricating microlens arrays by surface wrinkling. *Adv. Mater.* **18**, 3238-3242 (2006).
8. Chandra, D., Yang, S. & Lin, P. C. Strain responsive concave and convex microlens arrays. *Appl. Phys. Lett.* **91**, 3 (2007). https://doi.org/251912

Artn 251912

9. Khang, D. Y., Jiang, H. Q., Huang, Y. & Rogers, J. A. A stretchable form of single-crystal silicon for high-performance electronics on rubber substrates. *Science* **311**, 208-212 (2006).
10. Sun, Y. G., Choi, W. M., Jiang, H. Q., Huang, Y. G. Y. & Rogers, J. A. Controlled buckling of semiconductor nanoribbons for stretchable electronics. *Nat. Nanotechnol.* **1**, 201-207 (2006).
11. Sun, Y. G., Kumar, V., Adesida, I. & Rogers, J. A. Buckled and wavy ribbons of GaAs for high-performance electronics on elastomeric substrates. *Adv. Mater.* **18**, 2857-2862 (2006).
12. Kim, Y. H. *et al.* Spontaneous Surface Flattening via Layer-by-Layer Assembly of Interdiffusing Polyelectrolyte Multilayers. *Langmuir* **26**, 17756-17763 (2010). https://doi.org/10.1021/la103282m
13. Jiang, X. Y. *et al.* Controlling mammalian cell spreading and cytoskeletal arrangement with conveniently fabricated continuous wavy features on poly(dimethylsiloxane). *Langmuir* **18**, 3273-3280 (2002). https://doi.org/10.1021/la011668+
14. Efimenko, K., Finlay, J., Callow, M. E., Callow, J. A. & Genzer, J. Development and Testing of Hierarchically Wrinkled Coatings for Marine Antifouling. *ACS Appl. Mater. Interfaces* **1**, 1031-1040 (2009). https://doi.org/10.1021/am9000562
15. Bowden, N., Brittain, S., Evans, A. G., Hutchinson, J. W. & Whitesides, G. M. Spontaneous formation of ordered structures in thin films of metals supported on an elastomeric polymer. *Nature* **393**, 146-149 (1998).
16. Yoo, P. J., Suh, K. Y., Park, S. Y. & Lee, H. H. Physical self-assembly of microstructures by anisotropic buckling. *Adv. Mater.* **14**, 1383-1387 (2002). https://doi.org/10.1002/1521-4095(20021002)14:19<1383::aid-adma1383>3.0.co;2-d
17. Guvendiren, M., Yang, S. & Burdick, J. A. Swelling-Induced Surface Patterns in Hydrogels with Gradient Crosslinking Density. *Adv. Funct. Mater.* **19**, 3038-3045 (2009). https://doi.org/10.1002/adfm.200900622

18  Choi, W. M. *et al.* Biaxially stretchable "Wavy" silicon nanomembranes. *Nano Lett.* **7**, 1655-1663 (2007).
19  Jiang, H. Q., Sun, Y. G., Rogers, J. A. & Huang, Y. G. Mechanics of precisely controlled thin film buckling on elastomeric substrate. *Appl. Phys. Lett.* **90**, 3 (2007). https://doi.org/133119
Artn 133119
20  Lin, P. C. & Yang, S. Spontaneous formation of one-dimensional ripples in transit to highly ordered two-dimensional herringbone structures through sequential and unequal biaxial mechanical stretching. *Appl. Phys. Lett.* **90**, 241903 (2007). https://doi.org/10.1063/1.2743939
21  Yang, S., Khare, K. & Lin, P.-C. Harnessing Surface Wrinkle Patterns in Soft Matter. *Adv. Funct. Mater.* **20**, 2550–2564 (2010). https://doi.org/10.1002/adfm.201000034
22  Huang, J. *et al.* Capillary Wrinkling of Floating Thin Polymer Films. *Science* **317**, 650-653 (2007). https://doi.org/10.1126/science.1144616
23  Huang, J., Davidovitch, B., Santangelo, C. D., Russell, T. P. & Menon, N. Smooth Cascade of Wrinkles at the Edge of a Floating Elastic Film. *Phys. Rev. Lett.* **105**, 038302 (2010). https://doi.org/10.1103/PhysRevLett.105.038302
24  Evans, A. A., Cheung, E., Nyberg, K. D. & Rowat, A. C. Wrinkling of milk skin is mediated by evaporation. *Soft Matter* **13**, 1056-1062 (2017).
25  Box, F. *et al.* Dynamics of wrinkling in ultrathin elastic sheets. *Proceedings of the National Academy of Sciences* **116**, 20875-20880 (2019). https://doi.org/doi:10.1073/pnas.1905755116
26  Tobasco, I. *et al.* Exact solutions for the wrinkle patterns of confined elastic shells. *Nat. Phys.* **18**, 1099-1104 (2022). https://doi.org/10.1038/s41567-022-01672-2
27  Chen, C.-M., Reed, J. C. & Yang, S. Guided wrinkling in swollen, pre-patterned photoresist thin films with a crosslinking gradient. *Soft Matter* **9**, 11007-11013 (2013). https://doi.org/10.1039/C3SM51881G
28  Suh, K. Y., Seo, S. M., Yoo, P. J. & Lee, H. H. Formation of regular nanoscale undulations on a thin polymer film imprinted by a soft mold. *J. Chem. Phys.* **124** (2006). https://doi.org/024710
10.1063/1.2150211
29  Yoo, P. J. Fabrication of Complexly Patterned Wavy Structures Using Self-Organized Anisotropic Wrinkling. *Electron. Mater. Lett.* **7**, 17-23 (2011). https://doi.org/10.1007/s13391-011-0303-8
30  Hou, H., Yin, J. & Jiang, X. Smart Patterned Surface with Dynamic Wrinkles. *Acc. Chem. Res.* **52**, 1025-1035 (2019). https://doi.org/10.1021/acs.accounts.8b00623
31  Janssens, S. D. *et al.* Boundary curvature effect on the wrinkling of thin suspended films. *Appl. Phys. Lett.* **116**, 193702 (2020). https://doi.org/10.1063/5.0006164
32  Wang, T. *et al.* Curvature tunes wrinkling in shells. *Int. J. Eng. Sci.* **164**, 103490 (2021). https://doi.org/https://doi.org/10.1016/j.ijengsci.2021.103490
33  Yuan, W. *et al.* Photochemical Design for Diverse Controllable Patterns in Self‐Wrinkling Films. *Advanced Materials*, 2400849 (2024).
34  Li, T. *et al.* Hierarchical 3D patterns with dynamic wrinkles produced by a photocontrolled diels–alder reaction on the surface. *Advanced Materials* **32**, 1906712 (2020).
35  Ueno, M. *et al.* Practical Chemistry of Long-Lasting Bubbles. *World J. Chem. Ed.* **4**, 32-44 (2016). https://doi.org/10.12691/wjce-4-2-2
36  Géminard, J. C., Żywocinski, A., Caillier, F. & Oswald, P. Observation of negative line tensions from Plateau border regions in dry foam films. *Philosoph. Mag. Lett.* **84**, 199-204 (2004). https://doi.org/10.1080/09500830010001646707

37  Denkov, N. *et al.* Mechanism of formation of two-dimensional crystals from latex particles on substrates. *Langmuir* **8**, 3183-3190 (1992).
38  Deegan, R. D. *et al.* Capillary flow as the cause of ring stains from dried liquid drops. *Nature* **389**, 827-829 (1997).
39  Cerda, E., Ravi-Chandar, K. & Mahadevan, L. Wrinkling of an elastic sheet under tension. *Nature* **419**, 579-580 (2002). https://doi.org/10.1038/419579b
40  Huang, R. & Im, S. H. Dynamics of wrinkle growth and coarsening in stressed thin films. *Phys. Rev. E* **74**, 026214 (2006). https://doi.org/10.1103/PhysRevE.74.026214
41  Kodio, O., Griffiths, I. M. & Vella, D. Lubricated wrinkles: Imposed constraints affect the dynamics of wrinkle coarsening. *Phys. Rev. Fluids* **2**, 014202 (2017). https://doi.org/10.1103/PhysRevFluids.2.014202